\documentclass[aps,twocolumn,showpacs,preprintnumbers,prl,superscriptaddress]{revtex4}

\usepackage{color}
\usepackage{graphicx}
\usepackage{graphics}

\begin{document}

\title{Accurate ionization potential of gold anionic clusters from density functional theory and many-body perturbation theory}

\author{A. Tanwar}
\affiliation{Center for Biomolecular Nanotechnologies @UNILE, Istituto Italiano di Tecnologia, Via Barsanti, I-73010 Arnesano  (LE), Italy}
\author{E. Fabiano}
\affiliation{National Nanotechnology Laboratory (NNL), Istituto Nanoscienze-CNR, Via per Arnesano 16, I-73100 Lecce, Italy}
\author{P. E. Trevisanutto}
\affiliation{National Nanotechnology Laboratory (NNL), Istituto Nanoscienze-CNR, Via per Arnesano 16, I-73100 Lecce, Italy}
\author{L. Chiodo}
\affiliation{Center for Biomolecular Nanotechnologies @UNILE, Istituto Italiano di Tecnologia, Via Barsanti, I-73010 Arnesano  (LE), Italy}
\affiliation{European Theoretical Spectroscopy Facility - ETSF}
\author{F. Della Sala}
\affiliation{National Nanotechnology Laboratory (NNL), Istituto Nanoscienze-CNR, Via per Arnesano 16, I-73100 Lecce, Italy}
\affiliation{Center for Biomolecular Nanotechnologies @UNILE, Istituto Italiano di Tecnologia, Via Barsanti, I-73010 Arnesano  (LE), Italy}

\date{\today}

\begin{abstract}
We present a theoretical study of the ionization potential in 
small anionic gold clusters, using density functional theory, with and without 
exact-exchange, and many body perturbation theory, namely the 
G$_0$W$_0$ approach.
We find that G$_0$W$_0$ is the best approach and correctly describes 
the first ionization potential with an accuracy of about 0.1 eV.
\end{abstract}


\maketitle

\section{Introduction}
The peculiar properties exhibited by gold clusters made them 
a popular subject for physical chemistry studies and numerous 
applications in recent years,
\cite{schmidbaur90,daniel04,pyykko04,pyykko05,pyykko08,yam08,hutchings08,mohr_book}.
Unlike bulk gold, which is essentially chemically inert 
\cite{hammer95}, gold clusters possess edge atoms that have a lower 
coordination and thus a higher reactivity.
As a consequence, a close interplay between the structural and electronic 
properties of different Au nano-particles is observed
and the determination of clusters sizes and shapes has a fundamental 
importance for both basic studies and applications.

Presently, the atomic structure of gold clusters can be established by
a number of different experimental techniques \cite{schooss10} including 
photoelectron spectroscopy (PES) 
\cite{taylor92,hakkinen03,li03,bulusu06,yoon07,hakkinen08,shao10,huang10}, 
ion mobility \cite{furche02}, infrared spectroscopy \cite{gruene08,fielicke04},
trapped ion electron diffraction 
\cite{xing06,lechtken08,johansson08,lechtken09}, and
photodissociation \cite{schweizer03,gibb04,gloess08}. However, in all cases
detailed information from theoretical investigations must complement the
experimental data to allow the correct interpretation of the results and the
final resolution of the different isomers.
In this context, PES is a typical case, where photoelectron spectra
obtained from theoretical calculations on a set of plausible structures are
compared to the experimental peaks to assign the different features and 
identify the structure of the clusters.

Usually, the photoelectron spectrum of gold clusters is  
simulated by density functional theory (DFT) calculations, since
these are easily affordable for a large variety of structures and
provide direct access to approximate photoelectron spectra through
the inspection of single-particle orbital energies (within the
Kohn-Sham (KS) scheme \cite{ks}). However, generally the DFT calculations
are performed using the generalized gradient approximation (GGA) 
for the exchange-correlation (XC) functional, and thus several 
drawbacks are introduced. The most important in the present context 
is the limitation regarding the description of the single-particle spectrum 
by these functionals. In fact, correct values of the orbital 
energies cannot be achieved in DFT when 
an approximate description of the exchange interaction is employed
\cite{lhf,lundberg05,ls2}. 
Due to the self-interaction error (SIE) \cite{sie,sie2} 
present in GGA functionals,
orbital energies are strongly overestimated and for a comparison 
with experimental data a rescaling of the energies 
is required based on additional calculations
and/or experimental evidence \cite{hakkinen03,stowasser99}. 
Alternatively, for an accurate estimation of the
first ionization potential (IP), $\Delta$SCF calculations, based on total 
energy differences between the anionic and neutral species,
can be considered \cite{nagy00}.
 
More accurate orbital energies can be achieved with 
exact exchange (EXX) KS approaches, i.e. the 
optimized effective potential (OEP) method \cite{exx} 
or the localized Hartree-Fock (LHF) \cite{lhf} approximation. 
A recent work showed in fact that accurate photoelectron spectra can be
calculated for very small gold clusters (up to 4 atoms) by employing the LHF
method together with a Lee-Yang-Paar (LYP) \cite{lyp} correlation
functional \cite{fabiano09}. 
This approach is SIE-free and provides a good agreement
with reference experimental data without the need for a rescaling of the 
orbital energies. It is thus potentially a valuable tool for the simulation of 
photoelectron spectra and the interpretation of PES experiments. 
However, despite the LHFLYP approach is much more accurate 
than conventional GGA methods, it has still two drawbacks: 
i) the correlation part of the functional is treated at the GGA level, and 
ii) it is yet a KS method, thus based  on single-particle orbitals: 
even with the exact (unknown) XC functional the KS energy-level does not 
correspond to the exact PES levels \cite{chong02}, the only exception being 
the highest occupied molecular orbital (HOMO), which can be shown to be 
exactly related to the IP \cite{chong02,perdew82,perdew97}.
 
Alternative DFT approaches for the description of the PES levels 
are based on hybrid functionals,
where a fraction of the non-local Hartree-Fock exchange is 
included in the Hamiltonian.
This approach goes beyond the KS scheme and it is commonly 
applied for organic molecules.
However, it has several problems
when applied to metallic systems, because of the drawbacks of the
Hartree-Fock method in this case \cite{hpbeint,messmer82,post82}. 
Therefore its use is
essentially limited to small clusters. Moreover, hybrid approaches
are mostly based on an error cancellation effects and thus can neither fully 
correct self-interaction error nor can provide systematic improvements, while 
they introduce several additional parameters.
Reliable results can be obtained instead using range separated
hybrid functionals \cite {longRsep}, which take advantage of a
screening of the Hartree-Fock exchange by separating short- and long-range
effects.

The fundamental solution to probe PES experiments directly is to compute the 
quasiparticle energy spectrum \cite{onida02,duffy94}, 
as described by the quasiparticle equation
\begin{equation}\label{QPeq}
H_0(\mathbf{r})\phi^{QP}_i(\mathbf{r}) + \int dr'\Sigma_{xc}(\mathbf{r},\mathbf{r}',\epsilon^{QP}_i)\phi^{QP}_i(\mathbf{r}')
= \epsilon^{QP}_i\phi^{QP}_i(\mathbf{r})\ ,
\end{equation}
where $H_0=-(1/2)\nabla^2+v_{ext}+V_H$ is the single-particle Hamiltonian, 
$v_{ext}$ is the external (nuclear) potential, $V_H$ is
the Hartree potential and $\Sigma_{xc}$ is the self-energy term: 
a non-local, non Hermitian and energy dependent operator resulting 
from all electron many-body interactions. 
Here $\epsilon^{QP}_i$ and $\phi^{QP}_i$ are the quasiparticle (QP) energies 
and eigenstates, respectively. 
Unfortunately, calculations of the exact self energy for real 
systems are computationally infeasible and approximations are required.
We have calculated the self-energy operator within the 
GW approximation \cite{Hedin65,aryasetiawan98}.

\begin{equation}\label{e3GW}
\Sigma(\mathbf{r},\mathbf{r}',\omega) 
= \frac{i}{2\pi}\int e^{i\omega'0^{+}}G(\mathbf{r},\mathbf{r}',\omega-\omega')W(\mathbf{r},\mathbf{r}',\omega')d\omega'\ ,
\end{equation}
where  $G$ is the single particle Green's function and $W$ is 
the dynamically screened Coulomb potential.

In principle, Eq. (\ref{e3GW}) requires a self-consistent solution
\cite{pap1,pap2}. 
Nevertheless, for many systems the DFT wave functions are good 
approximations for the QP wavefunctions \cite{lhf,duffy94,chong02} and
the quasiparticle energies $\epsilon^{QP}_n$ can be obtained pertubatively 
("one-shot" or G$_0$W$_0$ calculations) from the 
starting one-electron energies $\epsilon^{0}_n$ as:
\begin{equation}\label{quasip}
\epsilon^{QP}_n=\epsilon^{0}_n+\Re  \left \langle \phi^{0}_{n} \right |\Sigma(\epsilon^{QP}_n)-v_{xc} \left | \phi^{0}_{n} \right\rangle\ . 
\end{equation}
Originally, G$_0$W$_0$ has been applied in solids and, recently, 
calculations on bulk gold showed a QP band structure 
in good agreement with experimental PES \cite{rang12}. Moreover, 
this approach has 
been used also for accurate calculations of the IP of finite-size systems, 
both in closed-shell 
\cite{shirley93,rohlfing00,gross01,ishii02,pavlyukh04,stan06,tiago06,kik06,noguchi08,tiago09,bruneval09,pap1,pap2,chiodo11,ke11,blase11,faber11,sharifzadeh12,Bruneval12,Umaril12,marom12,setten13}
and open-shell \cite{shirley93,lischner12} species.

\begin{figure}
\includegraphics[width=\columnwidth]{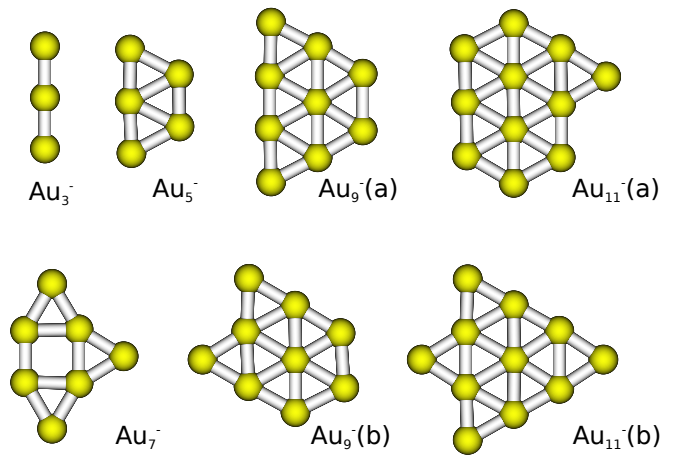}
\caption{\label{fig1} Structures of gold cluster anions}
\end{figure}

In this paper, we aim at assessing the merits and limitations of the
different methods illustrated above and perform a comparative study of
various computational approaches for the simulation of the first 
PES peak (i.e. IP) of several anionic gold clusters 
(up to 11 atoms; see fig. \ref{fig1}) 
for which experimental data are available. 

\section{Computational details}
Reference IP energies and the general structure of the clusters 
were taken from Ref. \cite{hakkinen03}. Geometries were
optimized using the PBEint XC functional \cite{pbeint},
which has been shown to yield accurate results for such 
systems \cite{pbeint_gold}.
For the larger clusters (Au$_7^-$, Au$_9^-$ and Au$_{11}^-$) we considered
the lowest two structures from Ref. \cite{hakkinen03}. However, 
for Au$_7^-$ at the PBEint/def2-TZVP level both isomers 
converged to the same geometry (Au$_7^-$ A in Ref. \cite{hakkinen03}). 
This was also confirmed by TPSS/def2-QZVPP calculations 
\cite{TPSS,tzvp}, which failed to identify a 
minimum-energy structure corresponding to Au$_7^-$ B. 
Thus, only one structure for Au$_7^-$ is considered here.
The IP of different systems was obtained from the HOMO energy of 
the following methods: Hartree-Fock (HF), LHF \cite{lhf}, LHFLYP
\cite{lhf,lyp}, PBEint \cite{pbeint}, PBE0 \cite{pbe0}.
In addition, we considered the $\Delta$ SCF approach (PBEint functional) 
and the post-DFT many-body perturbation theory G$_0$W$_0$ approach 
(from PBEint orbitals; G$_0$W$_0$@PBEint).
Finally, to asses the role of the reference orbitals for G$_0$W$_0$
calculations we performed our perturbative treatment also starting 
from self-consitent static GW one-electron energy and orbital results 
(G$_0$W$_0$@COHSEX), which have been recently shown to yield in
solid-state physics similar results as the more computationally expensive 
self-consistent GW calculations \cite{Bruneval06,Gatti07} and
good agrement with the experimental transport properties of
strongly correlated systems \cite{pap2,Gatti07,Tonatiuh11}
The COHSEX \cite{Bruneval06} is in fact a static version of the 
GW approximation, where the self-energy can be written as
\begin{eqnarray}\label{eSex}
\Sigma(\mathbf{r},\mathbf{r}')
&=&-\sum_{i}^{occ} \phi^{QP}_i(\mathbf{r})\phi^{QP}_i(\mathbf{r}')W(\mathbf{r},\mathbf{r}',\omega = 0) \nonumber \\
&+&\frac{1}{2}\delta(\mathbf{r}-\mathbf{r}')( W(\mathbf{r},\mathbf{r}',\omega = 0)-v(\mathbf{r}-\mathbf{r}') )
\end{eqnarray}
i.e. as a sum of an  instantaneous screened exchange (SEX) term and
a Coulomb hole (COH) correlation term.

All DFT calculations were performed with the TURBOMOLE program package
\cite{turbomole}, using the def2-TZVP basis set \cite{tzvp} and a scalar 
relativistic effective core potential (ECP) replacing the 
inner 60 electrons of each gold atom \cite{andrae90}.
The G$_0$W$_0$ and the self-consistent COHSEX calculations 
were performed by using the FHI-AIMS program \cite{aims1,aims2}
with the fourth-tier tight basis set (including additional $f$- and 
$g$-type functions) and scalar relativistic effects included 
through the zeroth-order relativistic approximation (ZORA) \cite{zora}.
With this approach we estimate the G$_0$W$_0$ orbital energies to be converged
within 0.05 eV. Test calculations for the PBE0 and PBEint functionals showed
that the results obtained by the two programs also differ by less than
0.05 eV.
In GW calculations the self energy $\Sigma$ was first calculated on
the immaginary frequency $\omega$-grid and then analytically continued 
to the real axis grid as implemented in FHI-AIMS code \cite{aimgen}.

\section{Results and discussion}
\begin{table*}
\begin{center}
\caption{ Ionization potential for several anionic gold clusters as resulting from different approaches: HF, LHF, LHFLYP,PBEint, $\Delta$SCF, PBE0, COHSEX, G$_0$W$_0$@COHSEX, G$_0$W$_0$@PBEint. Experimental data \cite{hakkinen03} are also reported. The last line shows the mean absolute error (MAE), excluding Au$_9^-$(b) and Au$_{11}^-$(a). All results are in eV.}
\label{tab2}
\begin{tabular}{lcccccccccc}\hline \hline
        &        &  &  & & & &      &\multicolumn{2}{c}{G$_0$W$_0$} & \\
\cline{9-10}
Cluster & HF & LHF & LHFLYP & PBEint & $\Delta$SCF & PBE0 & COHSEX & @COHSEX & @PBEint & Exp. \\
\hline 
Au$-$	    & 0.80 &  0.85 &  1.51 & -0.84 & 2.09 & -0.19 &  3.00  & 1.91  & 2.23 & 2.33 \\
Au$_3^-$    & 2.46 &  2.88 &  3.57 &  1.20 & 3.57 &  1.82 &  4.84  & 3.69  & 3.92 & 3.88 \\
Au$_5^-$    & 1.91 &  2.33 &  3.02 &  1.01 & 3.03 &  1.43 &  4.08  & 2.84  & 2.87 & 3.09 \\
Au$_7^-$    & 2.15 &  2.43 &  3.14 &  1.59 & 3.39 &  1.89 &  4.45  & 3.25  & 3.30 & 3.46 \\
Au$_9^-$(a) & 2.89 &  3.55 &  4.28 &  1.95 & 3.71 &  2.45 &  4.98  & 3.75  & 3.57 & 3.83 \\
Au$_9^-$(b)  & 2.36 &  2.83 &  3.56 &  1.73 & 3.44 &  2.09 &    -   &  -    & 3.30 & (3.83) \\
Au$_{11}^-$(a)& 2.69 &  3.23 &  3.96 &  2.03 & 3.65 &  2.39 &    -   &  -    & 3.52 & (3.80) \\         
Au$_{11}^-$(b)&3.05 &  3.77 &  4.50 &  2.18 & 3.81 &  2.62 &    -   &  -    & 3.73 & 3.80 \\         
            &      &       &       &       &      &       &        &       &      & \\
MAE         & 1.19 &  0.76 &  0.45 &  2.22 & 0.14 &  1.73 &  0.95  & 0.23  & 0.14 & \\ \hline \hline
\end{tabular}
\end{center}
\end{table*}

The calculated IPs, as obtained with different methods,
are listed in Tab. \ref{tab2}, together with the experimental results.
For Au$_9^-$ and  Au$_{11}^-$ we report two isomers, because they are
rather close in energy (at the PBEint level the (b) isomers are 0.21 and
0.07 eV higher for Au$_9^-$ and  Au$_{11}^-$, respectively).
For Au$_9^-$(b) and Au$_{11}^-$ COHSEX results are not present, because of the
excessive computational cost.
The last line of Tab. \ref{tab2} reports the mean absolute error (MAE)
with respect to experiment. This is calculated including all clusters but
Au$_9^-$(b) and Au$_{11}^-$(a), because for these two clusters our best estimates
($\Delta$SCF and G$_0$W$_0$@PBEint) show that the IP is significantly lower than
the experimental one, suggesting that these clusters are not measured in experiments.

The results confirm that the HOMO energy from GGA DFT calculations
(PBEint results) is a very poor estimation of the IP, because the 
SIE leads to a strong underestimation of the orbital energies.
Much better results are found instead when the Kohn-Sham calculations
are performed using an effective exact-exchange approach, as the LHF one.
In this case in fact the MAE drops to 0.66 eV and even further
improvement can be achieved by inclusion of explicit correlation
contributions (LHFLYP has a MAE of only 0.45 eV in line with the 
findings of Ref. \cite{fabiano09}). We note, nonetheless,  that
this improvement is not systematic, as shown by the fact that
LHFLYP performs well for the smaller clusters but it is worst than LHF for
the largest ones. A possible reason for such a behavior traces 
back to the limitations of the semilocal LYP correlation potential \cite{ls2} 
(similarly to other semilocal potentials). It is not fully compatible with
exact-exchange or OEP approaches \cite{loc}, and,
more important, fails to  properly describe
the correlation effects in extended systems of metallic character
\cite{paier07}. 
The net effect of the LYP functional is always to increase (in absolute value)
the HOMO energy by approximately 0.7 eV, independently on the system under
consideration. On the other hand, G$_0$W$_0$ calculations suggest that the
correlation effects are twice as big for the Au$^-$ atom as for the
largest clusters considered in this work, in agreement with the LHF vs.
experimental trend.

The importance of correlation effects is also remarked by a comparison
of the LHF results with the Hartree-Fock ones, as the latter display a larger
overall MAE. The difference can be in fact attributed not only to 
the local KS approximation (full-OEP exact-exchange Kohn-Sham 
results are often closer to HF than LHF ones \cite{lhf}), but also to
the LHF approximations to the full OEP, which make effective correlation 
contributions appear in the KS potential, so that an effective 
partial screening of the exchange is obtained \cite{fabiano09}. 
On the contrary,  in the Hartree-Fock method the correlation 
is completely lacking (absence of any screening).
Thus, Hartree-Fock calculations yield an overestimation 
of the electron-electron repulsion in metallic systems and the
IP obtained through the HOMO energy is underestimated (for the same reason the 
$s$-$d$ splitting is overestimated in Hartree-Fock calculations 
on metallic systems \cite{ijqc}). This effect is instead
mitigated in LHF calculations which in fact yield larger values for the IP.
Finally, poor results are also obtained by considering the HOMO
energy from hybrid calculations, as both GGA and HF results underestimate
the reference experimental values, so that no benefit from
error compensation can be obtained. 

It is thus clear that to achieve a HOMO energy of high accuracy
a proper treatment of exchange and correlation is necessary.
This is indeed obtained within the G$_0$W$_0$(@PBEint) approach
which in fact yields a MAE of only 0.14 eV, similarly with the 
$\Delta$SCF approach which is based on total energy differences (even 
slightly better G$_0$W$_0$ results are found using PBE starting orbitals, 
with a MAE of 0.11 eV; note however that the difference is below the 
estimated accuracy for G$_0$W$_0$ calculations).
We remark that the $\Delta$SCF method can be only applied for the
HOMO energy and few other orbital energies corresponding to the
highest of each irreducible symmetry representation. On the
contrary, the G$_0$W$_0$ can be applied straightforwardly for the
calculation of all the QP energies. We note however that
the perturbative G$_0$W$_0$ success in these calculations is
not trivial: this method, originally applied to periodic systems, 
has quite few studies about molecules and clusters
where the inhomogenity of the density and low coordination number 
of atoms play an important role.

In this sense it is interesting to observe that for the present gold cluster
G$_0$W$_0$ calculations on top of self-consistent 
COHSEX (GW@COHSEX) show rather worse results than the GW@PBEint ones,
in contrast to what observed in solid-state calculations \cite{Gatti07}.
This is because of the poor quality of the COHSEX orbitals 
which overestimate quite
significantly the reference ones. This behavior of the orbitals
may trace back to the fact that in all clusters the HOMO is 
essentially described by a linear combination of $s$ and $p$ atomic orbitals,
with negligible contribution from $d$ and $f$ atomic orbitals.
In fact, scf-COHSEX was demonstrated to be a good starting point for 
G$_0$W$_0$ when localized $d$ and $f$ orbitals \cite{Gatti07} are important 
(e.g., in strongly correlated systems). However, it can be expected 
to yield a poor description of delocalized electrons as those lying in
$s$ and $p$ orbitals, due to the lack of dynamical screening. 

Finally, it is worthwhile to note that for the present case of
gold anionic clusters the Hartree-Fock calculations always underestimate
the reference IP, whereas the G$_0$W$_0$ (and COHSEX) calculations
yield an increase (in absolute value) of the HOMO energy with respect
to HF. This behavior is opposite with respect to that usually observed
in molecular systems where HF overestimates the IP due to the 
too negative value of the non-local exchange and GW has the effect 
of reducing the HOMO energy thanks to the screening of the exchange 
\cite{pap1}. However, many exceptions exist, e.g. in atoms
\cite{shirley93,Bruneval12}, molecules \cite{gross01}
and metal clusters \cite{bruneval09,pap1,setten13}.
This latter behavior may be related to the important role of the
correlation in systems such as the gold clusters studied here, 
which overtakes the screening. In fact, COHSEX calculations for the 
Au$^-$ atom show that the effect of the screening term is to shift-up 
the HOMO energy by 0.2 eV, whereas the Coulomb hole shifts it down by 2.1 eV. 
The anionic character of the clusters instead appears to play a
negligible role, as similar findings are found also for neutral gold clusters
\cite{setten13}.

\section{Conclusions} 
In conclusion, we have shown that refined many body calculations 
at the G$_0$W$_0$@PBEint level can describe the ionization potential of 
gold clusters with an accuracy of about 0.1 eV, outperforming any other
orbital-based method. This result is a non-trivial confirmation
of the goodness of the GW approach for metallic clusters and shows that
the attainable accuracy is sufficient to distinguish 
between different isomeric forms of a cluster of a given size.
Thus the G$_0$W$_0$ approach can be used to select
specific clusters geometries by comparison with experimental
measurements (as in the case of Au$_9^-$ and Au$_{11}^-$).
Similar results can be achieved at the $\Delta$SCF level,
but losing any orbital picture.
Finally, within the Kohn-Sham scheme, relatively accurate results
can also be obtained when the effective exact exchange LHF method
is employed (eventually in conjunction with a semilocal correlation 
functional).
This approach has a considerably lower computational cost than the
G$_0$W$_0$  and thus may be considered as a cheaper alternative 
to the latter.

\section{Acknowledgments}
This work was partially funded by the European Research Council (ERC) 
Starting Grant FP7 Project DEDOM, grant agreement no. 207441.
We thank TURBOMOLE GmbH for providing the TURBOMOLE program package
and M. Margarito for technical support.


\begin{thebibliography}{99}
%
\bibitem{schmidbaur90} H. Schmidbaur, Gold Bulletin \textbf{23}, 11 (1990).
%
\bibitem{daniel04} M.-C. Daniel, D. Astruc, Chem. Rev. \textbf{104}, 293 (2004).
%
\bibitem{pyykko04} P. Pyykk\"o, Angew. Chem. Int. Ed. \textbf{43}, 4412 (2004).
%
\bibitem{pyykko05} P. Pyykk\"o, Inorganica Chimica Acta \textbf{358}, 4113 (2005).
%
\bibitem{pyykko08} P. Pyykk\"o, Chem. Soc. Rev. \textbf{37}, 1967 (2008).
%
\bibitem{yam08} W.-W. Yam, E. C.-C. Cheng, Chem. Soc. Rev. \textbf{37}, 1806 (2008).
%
\bibitem{hutchings08} G. J. Hutchings, M. Brust, H. Schmidbaur, Chem. Soc. Rev. \textbf{37}, 1759 (2008).
%
\bibitem{mohr_book} \textit{Gold Chemistry. Applications and Future Directions in the Life Sciences}, edited by F. Mohr (Wiley-VCH, Weinheim, 2009).
%
\bibitem{hammer95} B. Hammer, J. K. Norskov, Nature \textbf{376}, 238 (1995).
%
\bibitem{schooss10} D. Schoos, P. Weis, O. Hampe, M. M.  Kappes, Phil. Trans. R. Soc. A \textbf{368}, 1915 (2010).

\bibitem{taylor92} K. J. Taylor, C. L. Pettiettehall, O. Cheshnovsky, R. E. Smalley, J. Chem. Phys. \textbf{96}, 3319 (1992).
%
\bibitem{hakkinen03} H. Ha\"akkinen, B. Yoon, U. Landman, X. Li, H. J. Zhai, L. S. Wang, J. Phys. Chem. A \textbf{107}, 6168 (2003).
%
\bibitem{li03} J. Li, X. Li, H. J. Zhai, L. S. Wang, Science \textbf{299}, 864 (2003).
%
\bibitem{bulusu06} S. Bulusu, X. Li, L. S. Wang, X. C. Zeng, Proc. Natl. Acad. Sci. U.S.A. \textbf{103}, 8326 (2006).
%
\bibitem{yoon07} B. Yoon, P. Koskinen, B. Huber, O. Kostko, B. von Issendorff, H. Häkkinen, M. Moseler, U. Landman, ChemPhysChem \textbf{8}, 157 (2007).
%
\bibitem{hakkinen08} H. H\"akkinen, Chem. Rev. \textbf{37}, 1847 (2008).
%
\bibitem{shao10} N. Shao, W. Huang, Y. Gao, L.-M. Wang, X. Li, L. S. Wang, X. C. Zeng, J. Am. Chem. Soc. \textbf{132}, 6596 (2010).
%
\bibitem{huang10} W. Huang, R. Pal, L.-M. Wang, X. C. Zeng, L. S. Wang, J. Chem. Phys. \textbf{132}, 054305 (2010).
%
\bibitem{furche02} F. Furche, R. Ahlrichs, P. Weis, C. Jacob, S. Gilb, T. Bierweiler, M. M. Kappes, J. Chem. Phys. \textbf{117}, 6982 (2002).
%
\bibitem{gruene08} P. Gruene, D. M. Rayner, B. Redlich, A. F. G. van der Meer, J. T. Lyon, G. Meijer, A. Fielicke, Science \textbf{321}, 674 (2008).
%
\bibitem{fielicke04} A. Fielicke, A. Kirilyuk, C. Ratsch, J. Behler, M. Scheffler, G. von Helden, G. Meijer, Phys. Rev. Lett. \textbf{93}, 023401 (2004).
%
\bibitem{xing06} X. Xing, B. Yoon, U. Landman, J. H. Parks, Phys. Rev. B \textbf{74}, 165423 (2006).
%
\bibitem{lechtken08} A. Lechtken, C. Neiss, J. Stairs, D. Schoss, J. Chem. Phys. \textbf{129}, 154304 (2008).
%
\bibitem{johansson08} M. P. Johansson, A. Lechtken, D. Schooss, M. M. Kappes, F. Furche, Phys. Rev. A \textbf{77}, 053202 (2008).
%
\bibitem{lechtken09} A. Lechtken, C. Neiss, M. M. Kappes, D. Schooss, Phys. Chem. Chem. Phys. \textbf{11}, 4344 (2009).
%
\bibitem{schweizer03} A. Schweizer, J. M. Weber, S. Gilb, H. Schneider, D. Schooss, M. M. Kappes, J. Chem. Phys. \textbf{119}, 3699 (2003).
%
\bibitem{gibb04}  S. Gilb, K. Jacobsen, D. Schooss, F. Furche, R. Ahlrichs, M. M. Kappes, J. Chem. Phys. \textbf{121}, 4619 (1994).
%
\bibitem{gloess08}  A. N. Gloess, H. Schneider, J. M. Weber, M.  M. Kappes, J. Chem. Phys. \textbf{128}, 114312 (2008). 
%
\bibitem{ks} W. Kohn, L. J. Sham, Phys. Rev. \textbf{140}, A1133 (1965).
%
\bibitem{lhf} F. Della Sala, A. G\"orling, J. Chem. Phys. \textbf{115}, 5718 (2001).
%
\bibitem{lundberg05} M. Lundberg, P. E. M. Siegbahn, J. Chem. Phys. \textbf{122}, 224103 (2005).
%
\bibitem{ls2} E. Fabiano, F. Della Sala, J. Chem. Phys. \textbf{126}, 214102 (2007).
%
\bibitem{sie} A. J. Cohen, P. Mori-S\'anchez, W. Yang, Science \textbf{321}, 792 (2008).
%
\bibitem{sie2} F. Della Sala, Theor. Chem. Acc. \textbf{117}, 981 (2007).
%
\bibitem{stowasser99} R. Stowasser, R. Hoffmann, J. Am. Chem. Soc. \textbf{121}, 3414 (1999).
%
\bibitem{nagy00} A. Nagy, H. Adachi, J. Phys. B: At. Mol. Opt. Phys. \textbf{33}, L585 (2000).
%
\bibitem{exx} F. Della Sala in \textit{Orbital-Dependent Exact-Exchange Methods in Density Functional Theory and Chemical Modelling Applications and Theory} Vol.~7 (Royal Society of Chemistry, RSC Publishing, 2010), pp. 115--161.
%
\bibitem{lyp} C. Lee, W. Yang, R. G. Parr, Phys. Rev. B \textbf{37}, 785 (1988).
%
\bibitem{fabiano09} E. Fabiano, M. Piacenza, F. Della Sala, Phys. Chem. Chem. Phys. \textbf{11}, 9160 (2009).
%
\bibitem{chong02} D. P. Chong, O. V. Gritsenko, E. J. Baerends, J. Chem. Phys. \textbf{116}, 1760 (2002).
%
\bibitem{perdew82} J. P. Perdew, R. G. Parr, M. Levy, J. L. Balduz, Phys. Rev. Lett. \textbf{49}, 1691 (1982).
%
\bibitem{perdew97} J. P. Perdew, M. Levy, Phys. Rev. B \textbf{56}, 16021 (1997).
%
\bibitem{hpbeint} E. Fabiano, L. A. Constantin, F. Della Sala, Int. J. Quant. Chem. (2013), doi: 10.1002/qua.24042.
%
\bibitem{messmer82} R. P. Messmer, T. C. Caves, C. M. Kao, Chem. Phys. Lett. \textbf{90}, 296 (1982).
%
\bibitem{post82} D. Post, E. J. Baerends, Chem. Phys. Lett. \textbf{86}, 176 (1982).
%
\bibitem{longRsep} T. Stein, H. Eisenberg, L. Kronik, R. Baer, Phys Rev. Lett. \textbf{105}, 266802 (2010).
%
\bibitem{onida02} G. Onida, L. Reining, A. Rubio, Rev. Mod. Phys. \textbf{74}, 601 (2002).
%
\bibitem{duffy94} P. Duffy, D. P. Chong, M. E. Casida, D. R. Salahub, Phys. Rev. A \textbf{50}, 4707 (1994).
%
\bibitem{Hedin65} L. Hedin, Phys. Rev. \textbf{139}, A796 (1965).
%
\bibitem{aryasetiawan98} F. Aryasetiawan, O. Gunnarsson, Rep. Prog. Phys. \textbf{61}, 237 (1998).
%
\bibitem{rang12} T. Rangel, D. Kecik, P. E. Trevisanutto, G. M. Rignanese, H. V. Swygenhoven, V. Olevano, Phys. Rev. B \textbf{86}, 125125 (2012).
%
\bibitem{shirley93} {E. L. Shirley, R. M. Martin, Phys. Rev. B \textbf{47}, 15404 (1993).}
%
\bibitem{rohlfing00} {M. Rohlfing, Int. J. Quantum Chem. \textbf{80}, 807 (2000).}
%
\bibitem{gross01} {J. C. Grossman, M. Rohlfing, L. Mitas, S. G. Louie, M. L. Cohen, Phys. Rev. Lett. \textbf{86}, 472 (2001).}
%
\bibitem{ishii02} {S. Ishii, K. Ohno, Y. Kawazoe, S. G. Louie, Phys. Rev. B \textbf{65}, 245109 (2002).}
%
\bibitem{pavlyukh04} {Y. Pavlyukh, W. H\"ubner, Phys. Lett. A \textbf{327}, 241 (2004).}
%
\bibitem{stan06} {A. Stan, N. E. Dahlen, R. van Leeuwen, Europhys. Lett. \textbf{76}, 298 (2006).}
%
\bibitem{tiago06} {M. L. Tiago, J. R. Chelikowsky, Phys. Rev. B \textbf{73}, 205334 (2006).}
%
\bibitem{kik06} {E. Kikuchi, S. Ishii, K. Ohno, Phys. Rev. B \textbf{74}, 195410 (2006).}
%
\bibitem{noguchi08} {Y. Noguchi, S. Ishii, K. Ohno, T. Sasaki, J. Chem. Phys. \textbf{129}, 104104 (2008).}
%
\bibitem{tiago09} {M. L. Tiago, J. C. Idrobo, S. \"O\v g\"ut, J. Jellinek, J. R. Chelikowsky, Phys. Rev. B \textbf{79}, 155419 (2009).}
%
\bibitem{bruneval09} {F. Bruneval, Phys. Rev. Lett. \textbf{103} 176403, (2009).}
%
\bibitem{pap1} {C. Rostgaard, K. W. Jacobsen, K. S. Thygesen, Phys. Rev. B \textbf{81}, 085103 (2010).} 
%
\bibitem{pap2} {M. Strange, C. Rostgaard, H. H\"akkinen, K. S. Thygesen, Phys. Rev. B \textbf{83}, 115108 (2011).}
%
\bibitem{chiodo11} {L. Chiodo, M. Salazar, A. H. Romero, S. Laricchia, F. Della Sala, A. Rubio, J. Chem. Phys. \textbf{135}, 244704 (2011).}
%
\bibitem{ke11} {S. Ke, Phys. Rev. B \textbf{84}, 205415 (2011).}
%
\bibitem{blase11} {X. Blase, C. Attaccalite, V. Olevano, Phys. Rev. B \textbf{83}, 115103 (2011).}
%
\bibitem{faber11} C. Faber, C. Attaccalite, V. Olevano, E. Runge, X. Blase, Phys. Rev. B \textbf{83}, 115123 (2011).
%
\bibitem{sharifzadeh12}  S. Sharifzadeh, I. Tamblyn, P. Doak, P. T. Darancet, J. B. Neaton, Eur. Phys. J. B \textbf{85}, 323 (2012). 
%
\bibitem{Bruneval12} F. Bruneval, J. Chem Phys. \textbf{136}, 194107 (2012).
%
\bibitem{Umaril12} P. Umari, S. Fabris, J. Chem Phys. \textbf{136}, 174310 (2012).
%
\bibitem{marom12} N. Marom, F. Caruso, X. Ren, O. T. Hofmann, T. K\"orzd\"orfer,J. R. Chelikowsky, A. Rubio, M. Scheffler, P. Rinke, Phys. Rev. B \textbf{86}, 245227 (2012).
%
\bibitem{setten13} {M. J. van Setten, F. Weigend, F. Evers, J. Chem. Theor. Comput. \textbf{9}, 232 (2013).}
%
\bibitem{lischner12} J. Lischner, J. Deslippe, M. Jain, S. G. Louie, Phys. Rev. Lett. \textbf{109}, 036406 (2012).
%
\bibitem{pbeint} E. Fabiano, L. A. Constantin, F. Della Sala, Phys. Rev. B \textbf{82}, 113104 ({2010}).
%
\bibitem{pbeint_gold} E. Fabiano, L. A. Constantin, F. {Della Sala}, J. Chem. Phys. \textbf{134}, 194112 (2011).
%
\bibitem{TPSS} J. Tao, J. P. Perdew, V. N. Staroverov, G. E. Scuseria, Phys Rev. Lett. \textbf{91}, 146401 (2003).
%
\bibitem{tzvp} F. Weigend, F. Furche, R. Ahlrichs, J. Chem. Phys. \textbf{119}, 12753 (2003).
%
\bibitem{pbe0} C. Adamo, V. Barone, J. Chem. Phys. \textbf{110}, 6158 (1999).
%
\bibitem{Bruneval06} F. Bruneval, N. Vast, L. Reining, Phys. Rev. B \textbf{74}, 045102 (2006).
%
\bibitem{Gatti07} M. Gatti, F. Bruneval, V. Olevano, L. Reining, Phys. Rev. Lett. \textbf{99}, 266402 (2007).
%
\bibitem{Tonatiuh11} T. Rangel, A. Ferretti, P. E. Trevisanutto, V. Olevano, G.-M. Rignanese, Phys. Rev. B \textbf{84}, 045426 (2011).
%
\bibitem{turbomole} TURBOMOLE V6.4 2012, a development of {University of Karlsruhe} and  {Forschungszentrum Karlsruhe GmbH}, 1989-2007, {TURBOMOLE GmbH}, since 2007; available from {\tt http://www.turbomole.com}.
%
\bibitem{andrae90} D. Andrae, U. H\"aussermann, M. Dolg, H. Stoll, H. Preuss, Theor. Chim. Acta \textbf{77}, 123 (1990).
%
\bibitem{aims1} V. Blum, R. Gehrke, F. Hanke, P. Havu, V. Havu, X. Ren, K. Reuter, M. Scheffler, Comput. Phys. Commun. \textbf{180}, 2175 (2009).
%
\bibitem{aims2} V. Havu, V. Blum, P. Havu, M. Scheffler, J. Comput. Phys. \textbf{228}, 8367 (2009).
%
\bibitem{zora} S. Faas, J. G. Snijders, J. H. van Lenthe, E. van Lenthe, E. J. Baerends, Chem. Phys. Lett. \textbf{246}, 632 (1995).
%
\bibitem{aimgen} X. Ren, P. Rinke, V. Blum, J. Wieferink, A. Tkatchenko, A. Sanfilippo, K. Reuter, M. Scheffler, New Journal of Physics \textbf{14}, 053020 (2012).
%
\bibitem{loc}  L. A. Constantin, E. Fabiano, F. Della Sala, Phys. Rev. B \textbf{86}, 035130 (2012).
%
\bibitem{paier07} J. Paier, M. Marsman, G. Kresse, J. Chem. Phys. \textbf{127}, 024103 (2007). 
%
\bibitem{ijqc} F. {Della Sala}, E. Fabiano, S. Laricchia, S. D'Agostino, M. Piacenza, Int. J. Quant. Chem. \textbf{110}, 2162 (2010).
%

\end{thebibliography}
\end{document}